\documentclass{aastex63}

\usepackage{ulem}
\usepackage{amsmath}

\shorttitle{IAGB Stars II.}
\shortauthors{Freedman et al.}
\graphicspath{{./}{figures/}}

\begin{document}
\title{I-Band Asymptotic Giant Branch (IAGB) Stars:\\ II. A First Estimate of their Precision and a Differential Zero Point\footnote{Based on observations made with the NASA/ESA Hubble Space Telescope, obtained at the Space Telescope Science Institute, which is operated by the Association of Universities for Research in Astronomy, Inc., under NASA contract NAS 5-26555.}}

\author{Wendy~L.~Freedman}\affil{Department of Astronomy \& Astrophysics, University of Chicago, 5640 South Ellis Avenue, Chicago, IL 60637}\affiliation{Kavli Institute for Cosmological Physics, University of Chicago,  5640 South Ellis Avenue, Chicago, IL 60637}\correspondingauthor{Wendy L. Freedman}\email{wfreedman@uchicago.edu}
\author{Barry~F.~Madore}\affil{Observatories of the Carnegie Institution for Science 813 Santa Barbara St., Pasadena, CA~91101}\affil{Department of Astronomy \& Astrophysics, University of Chicago, 5640 South Ellis Avenue, Chicago, IL 60637}
\author{Taylor Hoyt}\affil{Department of Astronomy \& Astrophysics, University of Chicago, 5640 South Ellis Avenue, Chicago, IL 60637}
\author{In Sung~Jang}\affil{Department of Astronomy \& Astrophysics, University of Chicago, 5640 South Ellis Avenue, Chicago, IL 60637}\affiliation{Kavli Institute for Cosmological Physics, University of Chicago,  5640 South Ellis Avenue, Chicago, IL 60637}
\author{Abigail~J.~Lee}\affil{Department of Astronomy \& Astrophysics, University of Chicago, 5640 South Ellis Avenue, Chicago, IL 60637}\affiliation{Kavli Institute for Cosmological Physics, University of Chicago,  5640 South Ellis Avenue, Chicago, IL 60637}
\author{Kayla A. Owens}\affil{Department of Astronomy \& Astrophysics, University of Chicago, 5640 South Ellis Avenue, Chicago, IL 60637}

\begin{abstract}

Hubble Space Telescope (HST) observations of 92 galaxies that have a strong showing of I-band Asymptotic Giant Branch (IAGB) stars in their color-magnitude diagrams (CMDs), are used to measure the relative offset between the mean apparent I-band magnitudes of the IAGB population and the corresponding apparent I-band magnitudes of the TRGB as measured in the same frames (and CMDs) of those individual galaxies. This first exploratory, large-sample comparison is independent of any extinction (foreground or internal) that may be shared by these two populations. The marginalized luminosity functions used to determine the modal value of the {\it IAGB } population are well fit by a single, symmetric Gaussian. The difference in the two apparent magnitudes (in the sense IAGB minus TRGB) is -0.589~mag, with a combined standard deviation of $\pm$0.119~mag.
Adopting $M_I =$ -4.05~mag for the TRGB stars, the modal absolute magnitude of the {\it IAGB} is then calculated to be $M_I(IAGB) =$ -4.64~$\pm$0.12~mag. The ensemble dispersion quoted above gives a standard error on the mean of $\pm$0.012~mag (based on the full sample of 92 galaxies). 
Independently, the three geometry-based zero points for I-band AGB stars are found (in Paper I) to be $M_I = $ -4.49 $\pm$0.003~mag in the LMC (4204 stars), $M_I = $ -4.67 $\pm$0.008~mag, for the SMC (916 stars) and $M_I = $ -4.78 $\pm$0.030~mag for NGC~4258 (62 stars), leading to a global zero-point (weighted) average of  $<M_I> = $ -4.64 $\pm$ 0.15~mag (stat). The scatter found in the anchors is comparable to the scatter in the field sample discussed here, but the calibration sample is small.
The application of this method to galaxies well outside of the Local Group, shows that these standard candles can readily be found and measured out to at least 9~Mpc, using already available archival data. 
\end{abstract}

\section{Introduction}

In Madore et al. (2024; hereafter Paper I), we introduced {\it I-band AGB} ({\it IAGB}) stars as possible extragalactic distance indicators. These stars are closely related, but not identical to the {\it J-Branch AGB} ({\it JAGB}) stars of Weinberg \& Nikolaev (2001), who drew attention to those stars as precision distance indicators, based on their near-infrared colors and magnitudes, as measured in the Large Magellanic Cloud (LMC). {\it JAGB} stars have subsequently been found to be excellent distance indicators in their own right (well beyond the LMC, see Freedman \& Madore 2023, Lee 2023), and they are now a core part of the {\it Chicago Carnegie Hubble Program} (CCHP) using Cepheids, TRGB stars and now {\it JAGB} stars to determine three independent distances to each of the host galaxies being used to calibrate the absolute magnitudes of Type Ia supernovae (Freedman 2021) en route to the Hubble constant. 
{ As discussed in	more detail in the Introduction to Paper I, {\it
IAGB} stars are defined by their extremely red colors in the I vs
(V-I) color-magnitude diagram (CMD). The blue cut-off is chosen to
maximize the number of IAGB stars to the red contributing to the
measured luminosity function, while simultaneously minimizing
contamination by the oxygen-rich population of precursor AGB stars
found to the blue. In order to illustrate the effect of perturbing
this blue cut-off around a nominal value, the three resulting luminosity
functions are shown color-coded in the far right panels of Figures 3 through 10, below. It is our
expectation that IAGB stars and JAGB stars are drawn from one and the
same parent population, differing only in the contribution of (the
bluest) stars still making the evolutionary transition from oxygen-rich to
carbon-rich stars.}

{\it IAGB} stars show the same uniquely identifiable colors and well-defined mean magnitudes as the {\it JAGB} stars, but the {\it IAGB} stars are defined at shorter wavelengths and bluer colors. This will prove to be an advantage for their discovery and application when {\it JWST} is considered, given the tighter point spread function (PSF) and finer angular resolution of the bluer filters available in the near-infrared imager, NIRCAM, on board {\it JWST}. NIRCAM's F090W filter is a reasonably close match to the I-band (F814W) discussed here.
In the application of the {\it JAGB} or {\it IAGB} methods at extreme distances, crowding in the outer disks will become a limiting factor, at which point the {\it IAGB} method will  prevail.

\section{ABSOLUTE CALIBRATION OF THE IAGB METHOD}

In Paper I of this program we used the geometric distances to the LMC, SMC and NGC~4258 to provide three independent estimates of the zero point of the {\it IAGB} Method. Taken as an ensemble, these three achor galaxies give $M_I$ = -4.64 $\pm$ 0.15~mag (stat). Each of these calibrating galaxies has its own strengths and weaknesses. In combination, their inter-comparison may be informative as to the impact of their differing systematics and/or reveal errors in the adopted assumptions used in calibrating each of these datasets. 


\medskip
As pointed out in Paper I, to first order there is no measurable trend of the {\it IAGB} luminosity with metallicity, given that the {\it IAGB} zero points for the LMC and NGC~4258 (which have intermediate metallicities of 8.50 and 9.06 dex, respectively) fall on either side of the zero point derived for the SMC, which has the lowest metallicity (12 + log(O/H) = 7.98 dex)\footnote{Metallicities taken from the recent compilation of Madore \& Freedman (2024) and references therein.} of the three. 
In Section 4 of this paper we offer a first look at the possible level of influence of star formation history on the mode of the {\it IAGB} luminosity function, finding no evidence so far.

We now  examine not only the zero point of the {\it IAGB} method, but also its precision and accuracy, including uncertainties due to sample size, differential reddening, variations in star formation history, regional crowding and metallicity. The scatter in Figure 1 puts a stringent upper limit on these ``second parameters." 

\section{A Differential Calibration of the IAGB Zero Point and its Dispersion}

Here we undertake the differential comparison of the TRGB and I-band AGB (IAGB) methods, using a homogeneously reduced and self-consistently analyzed set of HST images for 92 relatively nearby, highly resolved galaxies. This sample consists of galaxies that have well-established {\it TRGB} distances and also show sufficient numbers of {\it IAGB} stars for the construction of robust I-band {\it AGB} luminosity functions.  

The individual color-magnitude diagrams (CMD) and the resulting {\it IAGB} luminosity functions are shown in Figures 3 through 12. All of the data were retrieved from the {\it Extragalactic Distance Database} (EDD: Tully et al. 2009 \url{http:edd.ifa.hawaii.edu}). The entire collection of HST CMDs, for which EDD astronomers were able to obtain {\it TRGB} detections and measurements, were inspected by authors of this paper before special processing of the {\it IAGB} population. It needs to be noted that while all of the galaxies selected have I-band data (more specifically, [F814W] HST flight magnitude data) they do not all have data taken through the same blue-band filters with which a color is being formed. However, this is of little significance other than visibly shifting the blue color cut-off being adopted for a given galaxy, and used to define the redward {\it IAGB} sample.  Given an adopted blue cut-off, stars redder than that value have then been projected onto the vertical I-band axis, binned and smoothed to form the continuous luminosity function seen in black in the right half of each panel.  {\bf To aid the reader, in judging the sensitivity of the luminosity function to changing the color cut-off, we also plot in red and blue, respectively, the luminosity functions formed after moving the color cut-off by $\pm$0.1 mag around the nominal value.}
Because of the increasing density of stars approaching the ``vertical'' (O-type) AGB population situated to the blue, the red-colored luminosity function (having the largest amplitude) corresponds to the bluer cut-off (and the lower-amplitude, flanking luminosity function corresponds to the redder color cut-off). In most cases the modal value is stable to this level of variation in the method, but they are all shown  on a galaxy by galaxy basis.

\begin{figure*}
\centering
\includegraphics[width=1.2\textwidth]{"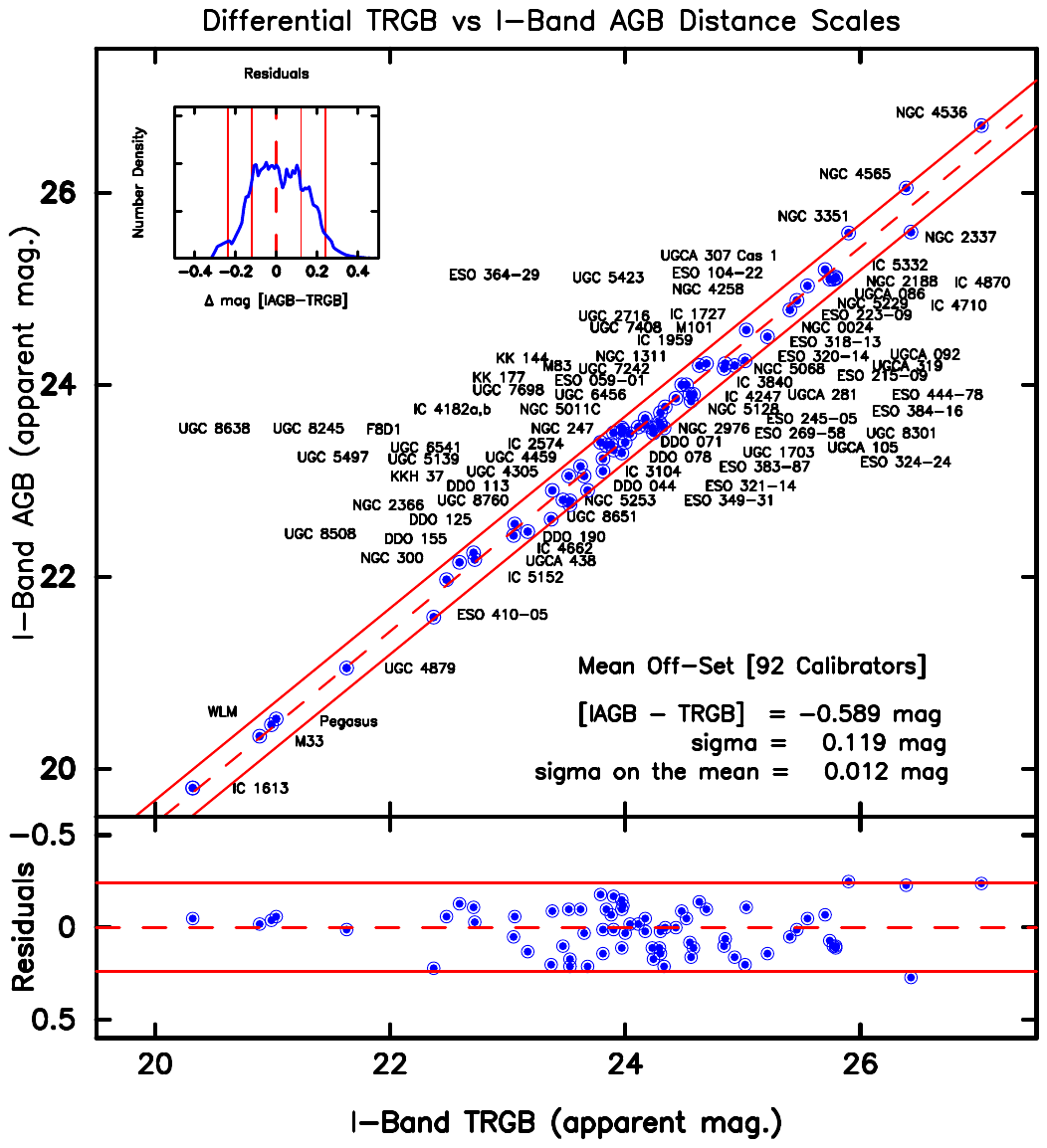"}
\caption{A differential comparison of the {\it TRGB} and {\it IAGB} distance scales. The apparent magnitude of the {\it TRGB} discontinuity is plotted on the horizontal axis. The apparent magnitude of the mode of the {\it IAGB} luminosity function is plotted on the vertical axis. The red dashed line has a slope of unity and has been fit to minimize the scatter of the data points around it. This minimization occurs at a value of -0.589 $\pm$ 0.003~mag (sigma on the mean) in the sense that the {\it IAGB} stars are brighter than the level of the TRGB. The measured inter-method scatter is $\pm$0.119~mag. The solid red lines are $\pm$ two sigma bounds. The residuals are plotted as a function of the {\it IAGB} apparent I-band magnitude the lower panel. 
The marginalized histogram of residuals is shown in the small inset found in the upper left corner of the main plot. 
}
\label{fig:f1}
\end{figure*}

In the right panels of Figures 3 through 14, a symmetric Gaussian (in blue) is then fit to the resulting luminosity function (in black) formed at the fiducial color cut-off. The fit is constrained by the peak in the observed luminosity function and by both of the wings of the distribution within 1-2 sigma of the peak. In most cases the fitted Gaussian underestimates the extended wings, which by choosing the mode, impose no bias one way or the other on the solution. That is, we are, in essence, simply measuring the dominant peak in the luminosity function, detached from the more broadly distributed background population. 
The sigmas on the fitted Gaussians range from 0.12 to 0.45 magnitudes. Using the fits to the LMC, SMC and NGC~4258, as given in Paper I, we found that the width of the (single-epoch) IAGB Gaussian distribution to be $\pm$0.32 mag. 
Individual dispersions are given in Column 4 of {\bf Table 1}, and may be useful in determining second-order correlations to the derived distance modulating especially if star-formation history, in the form of significant bursts, are of importance in specific galaxies. That possibility is explored later in this paper.

Tables 6 \&  7 contain the following information: [1] Galaxy Name\footnote{We note that some of the images used in EDD for TRGB tip detections were primarily intended for the study of disk populations. These galaxies are marked with an asterisk in this table, and the impact of their removal is discussed further in the Appendix}, [2] the TRGB apparent I-band magnitude, [3] the IAGB I-band modal magnitude followed by [4] the one-sigma width of the fitted Gaussian, [5] the number of {\it IAGB} stars within $\pm$ 2-sigma of the mode, [6] the sigma on the mean, [7] $V_{VGS}$ the expansion velocity corrected for Virgo Cluster, Great Attractor and Shapley Supercluster flow perturbations (NED 2023), [8] $A_I$ the I-band foreground Milky Way extinction (NED 2023), and  finally pairs of [8] \& [10] true distance moduli and [9] \& [11] distances for both the {\it TRGB} and {\it IAGB} measurements, assuming $M_I$ = -4.05 and -4.64~mag for the {\it TRGB} and {\it IAGB}, respectively.

Figure 1 plots the main result of this paper derived from Table 1. In the figure we give the apparent magnitude of tip of the red giant branch on the horizontal axis, and the apparent mean magnitude of the {\it IAGB} on the vertical axis.  This choice is made so as to make the plot independent of any foreground reddening or any {\it in situ} extinction that is shared by the two populations, along the same line of sight to the same host galaxy. The run of these two apparent-magnitude values is shown fit by unit slope line, having a zero point that is the mean offset between the two distance indicators. This mean offset (in the sense {\it TRGB} minus {\it IAGB}) is -0.589 magnitudes with a scatter of $\pm$ 0.119 mag. The two flanking lines, in the upper plot and in the lower residual plot, are $\pm$2 sigma from the unit slope line. Given 92 galaxies in the fit, we calculate an error on the mean of $0.119/\sqrt{91} = 0.012$~mag.  The distribution of residuals is shown as a small inset in the upper left corner of the main panel. 

By adopting the independently-determined absolute magnitude for the {\it TRGB} of $M_I(TRGB) = $ -4.05 $\pm 0.035$~(sys.)~mag (Freedman 2021), we can then derive the absolute magnitude for the {\it IAGB} stars, which can be compared to the values derived in Paper I. We find $M_o (IAGB) = $ -4.64 $\pm$ 0.012~mag, based on 92 galaxies. The value from the independent zero-point analysis in Paper~I is identical: $M_o (IAGB) = $ -4.64 $\pm$ 0.116~mag, albeit with lower statistical significance, being based on only three galaxies, whose individual deviations (LMC: +0.156~mag, SMC: -0.023~mag and N4258: -0.133~mag) all fall well within the 2-sigma ($\pm$ 0.24~mag) bounds of the far-field sample, shown at the bottom of Figure 1. 
\section{Nuisance Parameters}

It can be asserted  that, above and beyond the random noise from measurement errors, the effects of any  additional (non-covariant) parameters that can contribute to the scatter seen in Figure 1, will be collectively bounded by the observed scatter of $\sigma_I = \pm$ 0.12 ~mag, reported for the 92 galaxies so far observed.

A strong contribution to the observed scatter due to metallicity has already been tested for (and not found) using the geometric calibrators discussed in Paper~I. That test had only three data points, but each of those galaxies have distinctly different metallicities; moreover they each have accurate geometric distances. No trend of the {\it IAGB} zero point with metallicity was found.

Here we test for a possible effect of star formation history on the {\it IAGB}, but, again see no correlation.  We have already noted that the width of the Gaussian fit to individual {\it IAGB} luminosity functions varies over a considerable range of nearly a factor four: from 0.12 to 0.45~mag. It is conceivable that detailed differences in the star formation history 
over the time period producing the intermediate-age stars that are now populating the oxygen-rich AGB (which are the presumed progenitors of the {\it IAGB}) could influence the relative numbers of stars at distinct epochs, through bursts or pauses in the star formation rates, revealing themselves in the numbers of intermediate mass stars now entering advanced stage of evolution.
Such an activity-modulated weighting of the relative numbers of stars in the oxygen-rich and/or, possibly more importantly, the carbon-rich phase, could result in wider or narrower {\it IAGB} luminosity functions. Figure 2 attempts to test this possibility.

In Figure 2 we have taken the difference between the true {\it IAGB} distance modulus of a given galaxy and its true {\it TRGB} distance modulus, and plotted that difference as a function of that galaxy's {\it IAGB} luminosity function width. To the eye there in no discernible correlation, as a least squares fit (barely distinguishable from zero in the plot) and its one-sigma bounding slopes (broken lines) around zero, confirm. 

Finally, if all of the scatter were to be attributed, say, to differences in internal extinction between galaxies, then this would require $\sigma (A_I) = \sigma (IAGB) = \pm$~0.12~mag. This would translate into $\sigma(E(B-V) = \pm$0.07~mag, which is not beyond the realm of possibility. Comparing {\it IAGB} moduli with {\it JAGB} moduli for the same galaxies might reveal differential reddening's role here, given that the J-band extinction is a factor of two smaller than extinction measured in the I band.  We await {\it JAGB} distances to significantly more nearby galaxies; HST is fully capable of undertaking this task.

\begin{figure}
\centering
\includegraphics[width=15.9cm, angle=-0]{"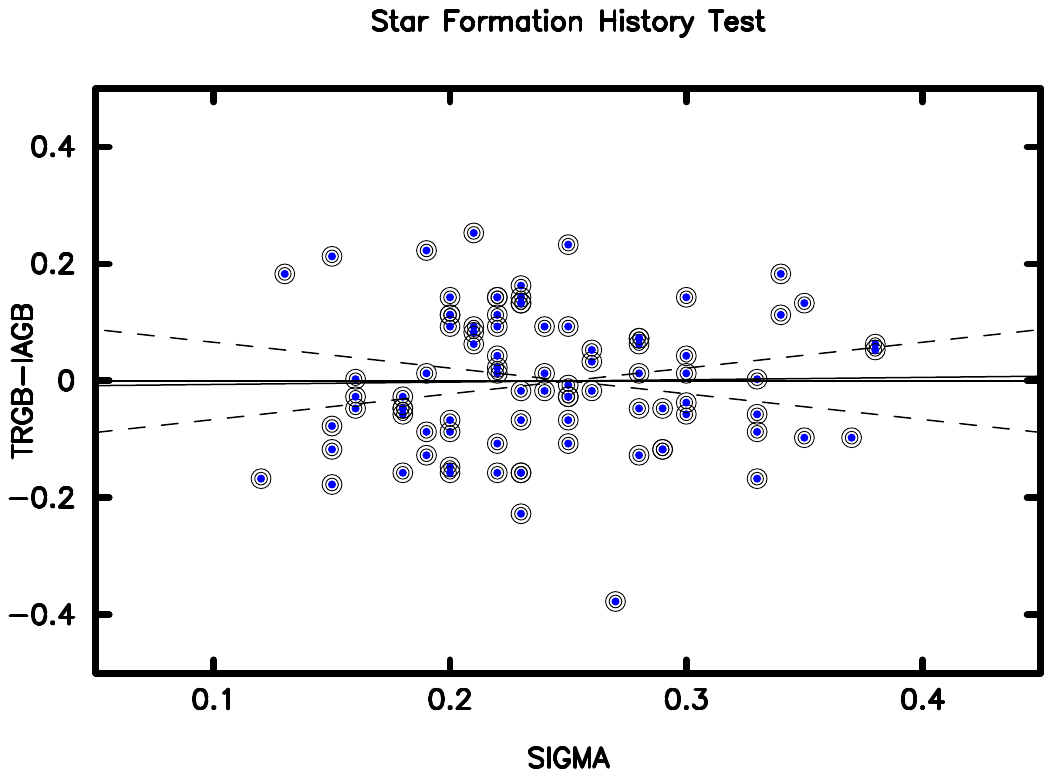"}
\label{fig:f9}
\caption{ Differences between the {\it TRGB} and {\it IAGB} true distance moduli as a function of the width of the {\it IAGB} luminosity function (expressed as a sigma). Broken lines are the one-sigma uncertainties on the statistically flat slope. The lack of a correlation with sigma suggests that star formation history over the age interval sampled by the {\it IAGB} stars is not perceptibly imprinted on the mode of the {\it IAGB} luminosity function.}
\end{figure}


\begin{figure}
\centering
\includegraphics[width=18.0cm, angle=-0]{"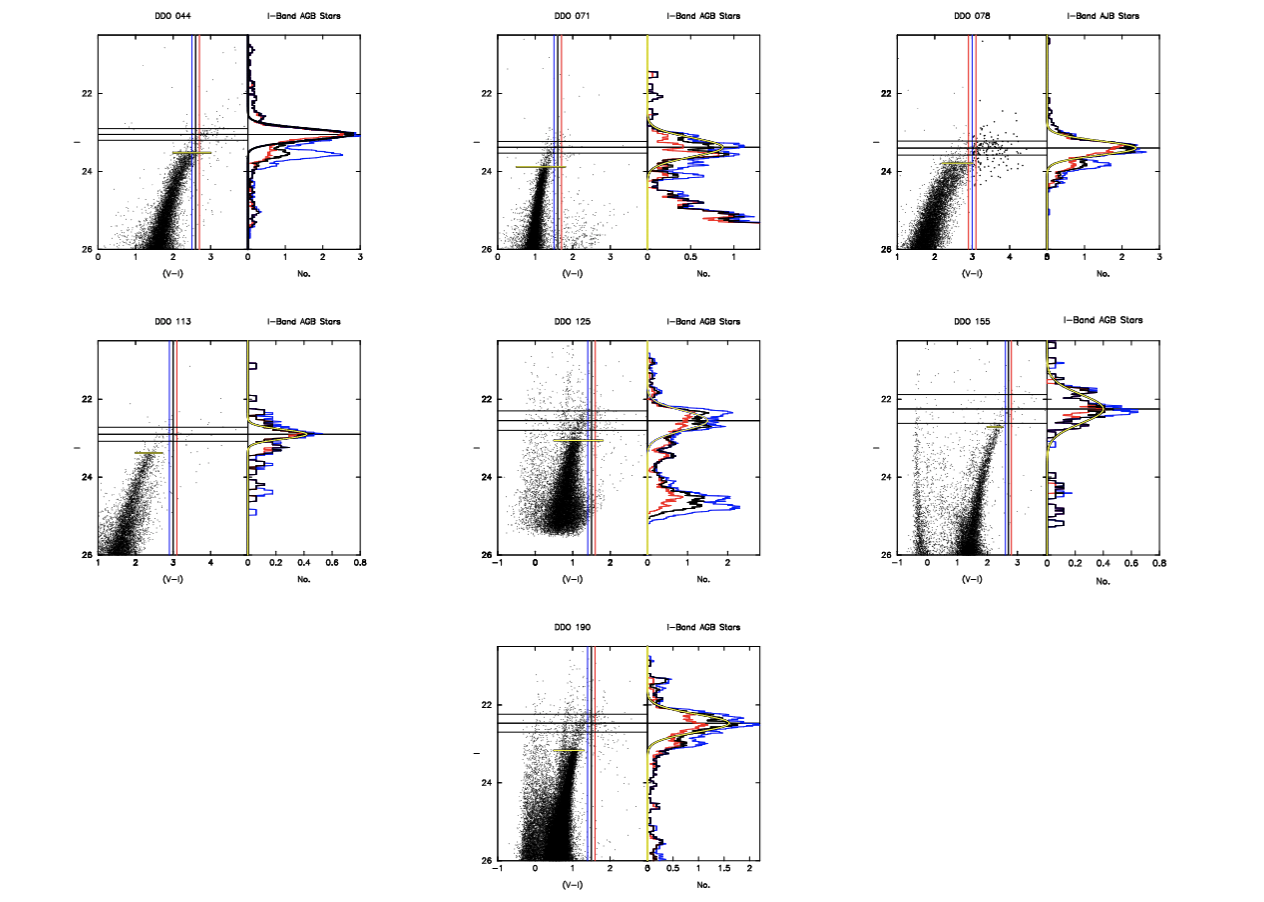"}
\caption{DDO Galaxies: I-Band AGB color-magnitude diagrams (left half of each panel) and I-band (F814W) luminosity functions (right half of each panel). { Horizontal black lines in the left panel mark the mode of the luminosity function (center line), flanked by lines at $\pm$ one-sigma in the Gaussian fit to the smoothed luminosity function. A shorter, yellow-on-black line marks the measured level of the TRGB. Thick (smooth) black lines in the right panels are symmetric Gaussian fits to the peak and wings of the smoothed/main luminosity function, also shown in black. Blue and Red luminosity functions correspond to color cuts that are shown in the CMD (left panel) on either side of the adopted color cut (blue vertical line).  Blue luminosity functions use the bluer color cut (larger numbers) and the red luminosity functions uses the redder color cut (smaller numbers).}  See text for additional details.}
\label{fig:f2}
\end{figure}

\begin{figure}
\centering
\includegraphics[width=18.0cm, angle=-0]{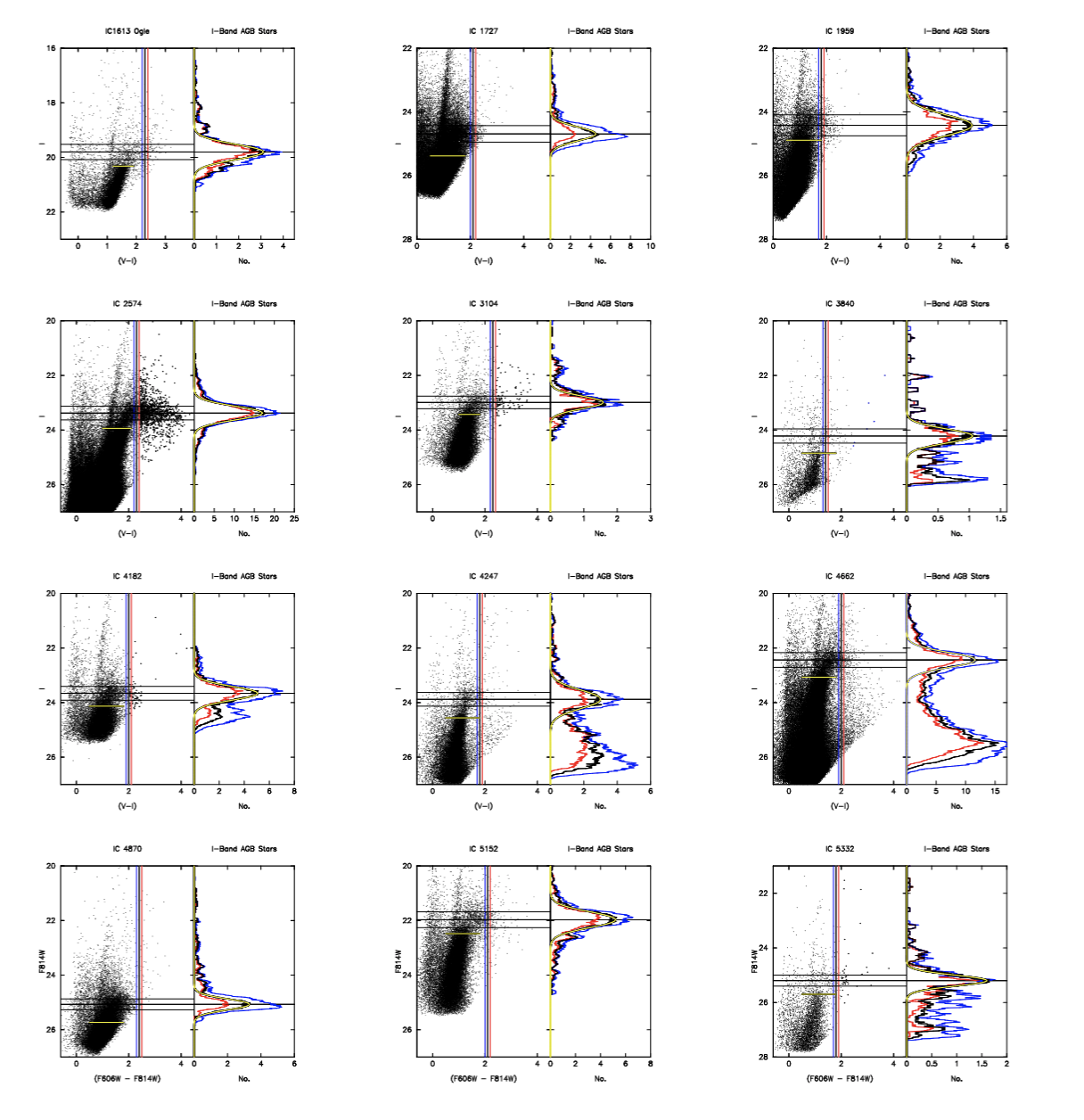}
\caption{IC galaxies (cont.) See Figure 3 for description.}
\end{figure}

\begin{figure}
\centering \includegraphics[width=18.0cm, angle=-0]{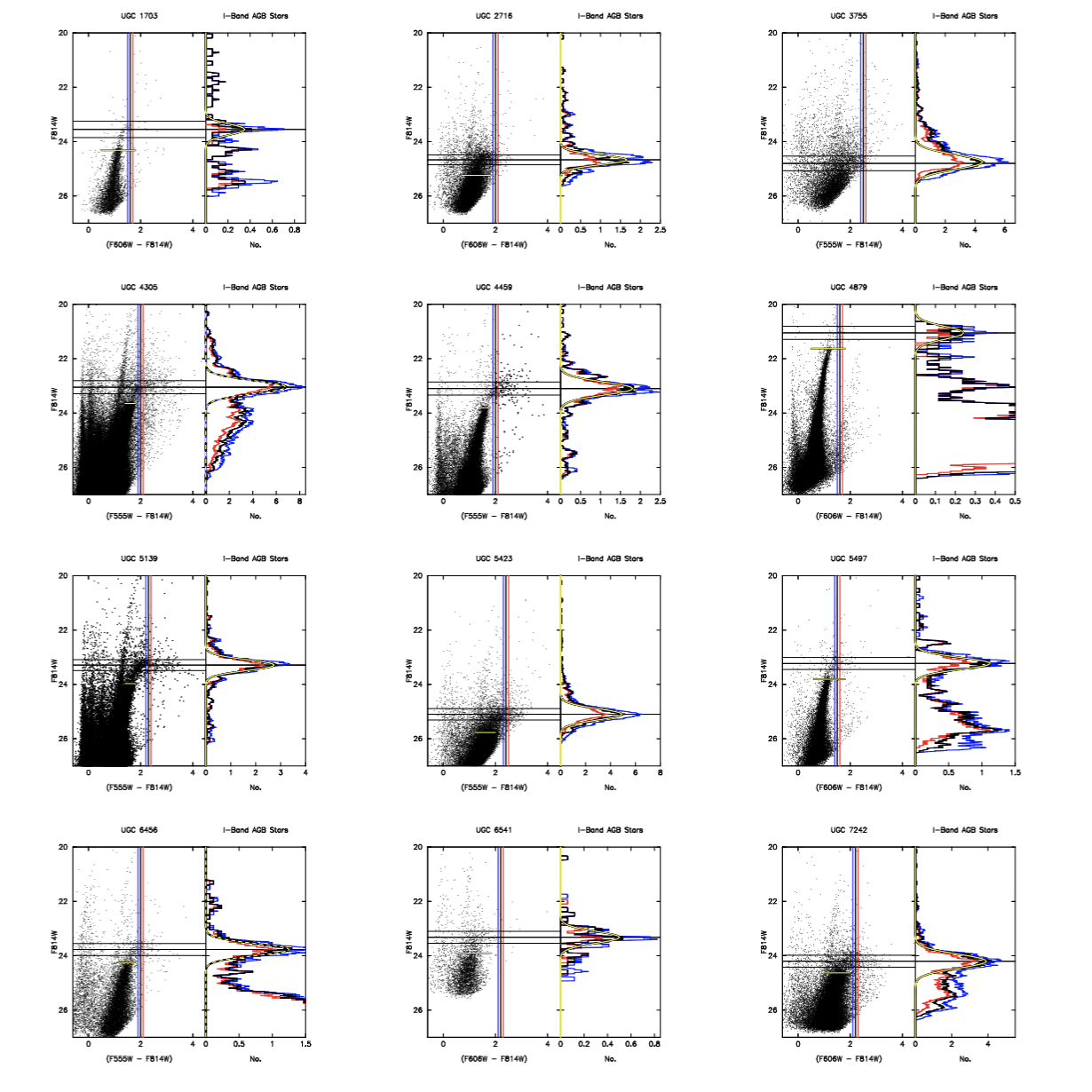}
\caption{UGC Galaxies. See Figure 3 for description.}
\end{figure}

\begin{figure}
\centering
\includegraphics[width=18.0cm, angle=-0]{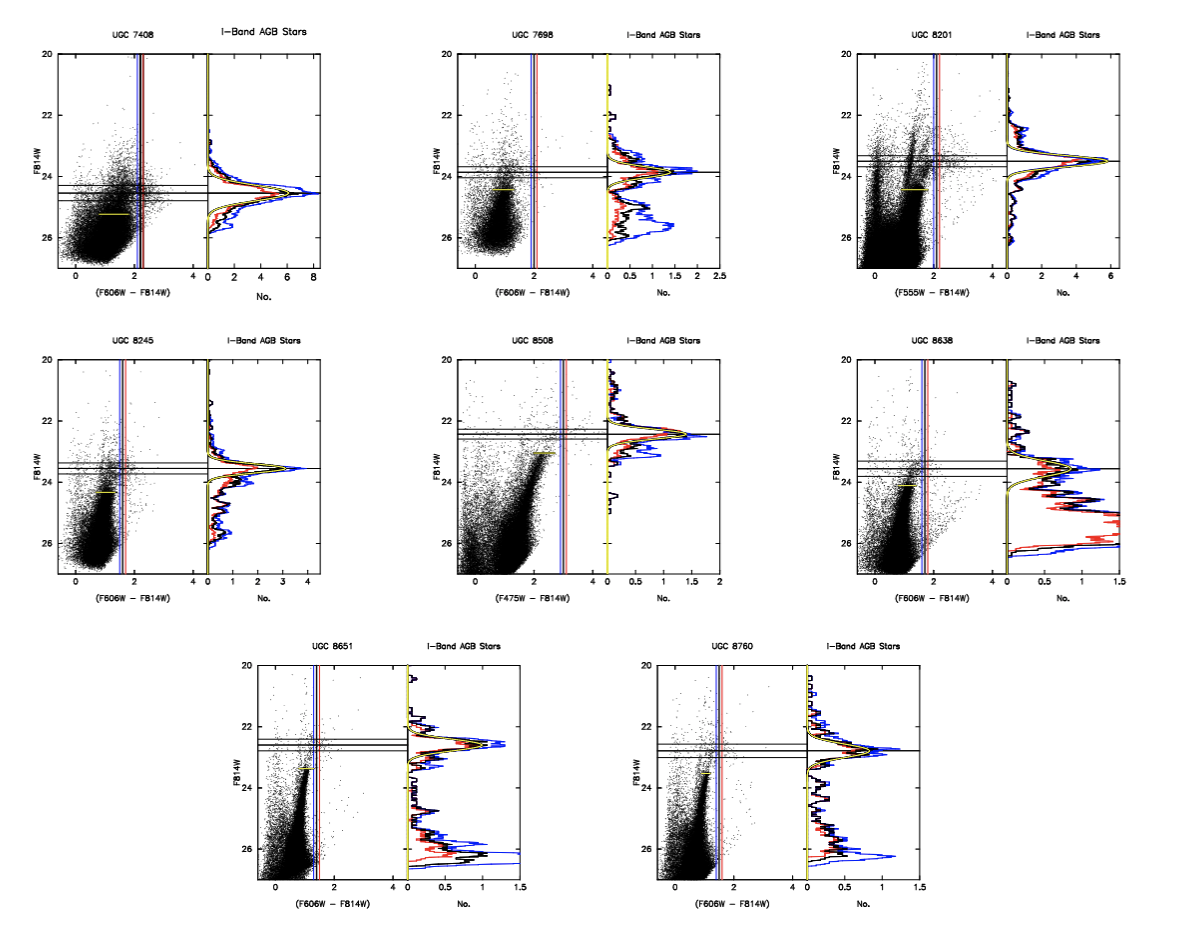}
\caption{UGC Galaxies (cont.) See Figure 3 for description.}
\end{figure}

\begin{figure}
\centering
\includegraphics[width=18.0cm, angle=-0]{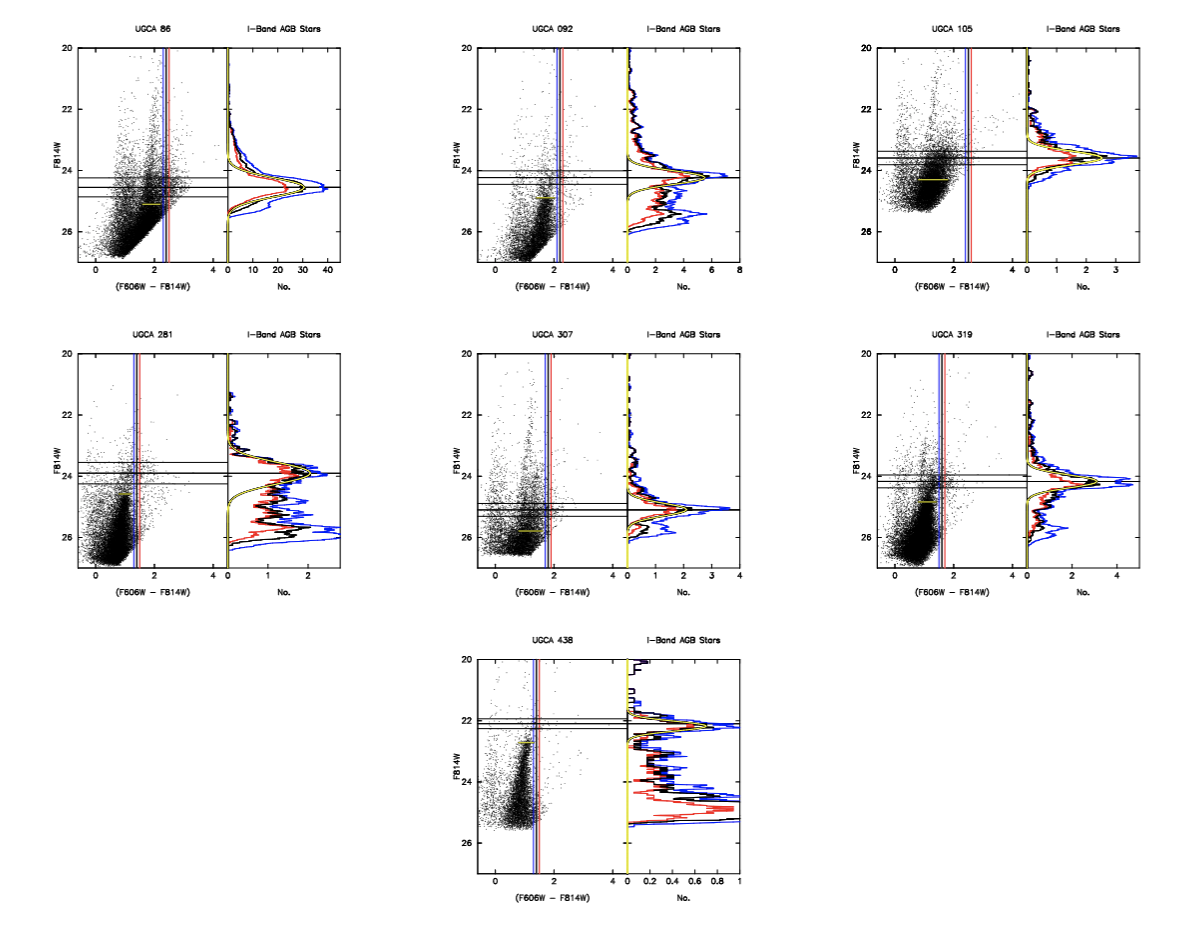}
\caption{UGCA galaxies. See Figure 3 for description.}
\end{figure}

\begin{figure}
\centering
\includegraphics[width=18.0cm, angle=-0]{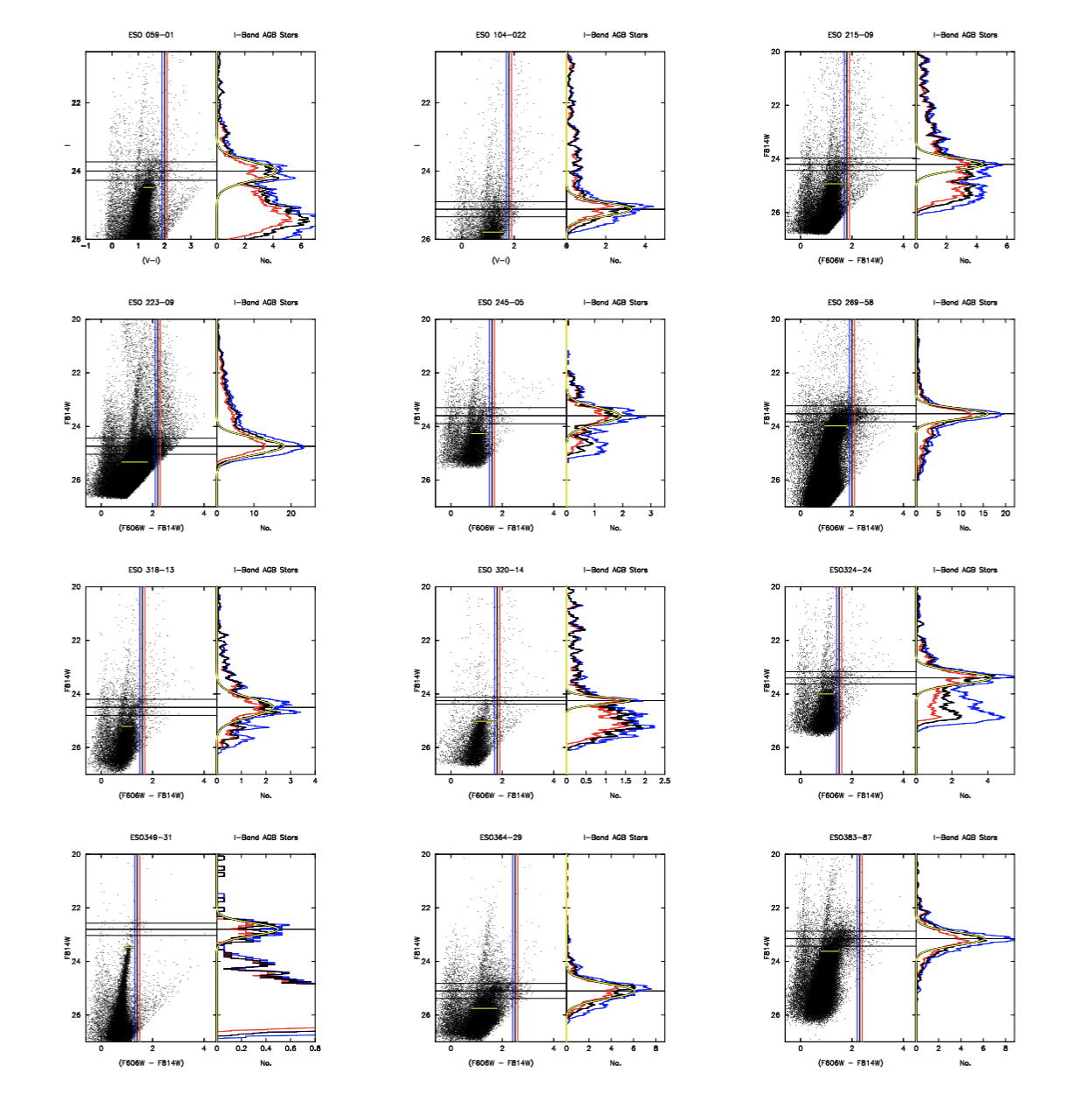}
\caption{ESO Galaxies. See Figure 3 for description.}
\end{figure}

\begin{figure}
\centering
\includegraphics[width=18.0cm, angle=-0]{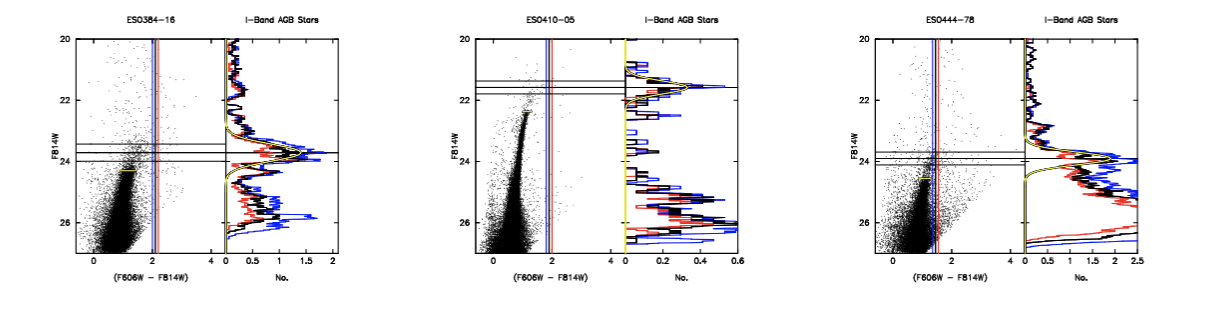}
\caption{ESO Galaxies (cont.) See Figure 2 for description.}
\end{figure}

\begin{figure}
\centering
\includegraphics[width=18.0cm, angle=-0]{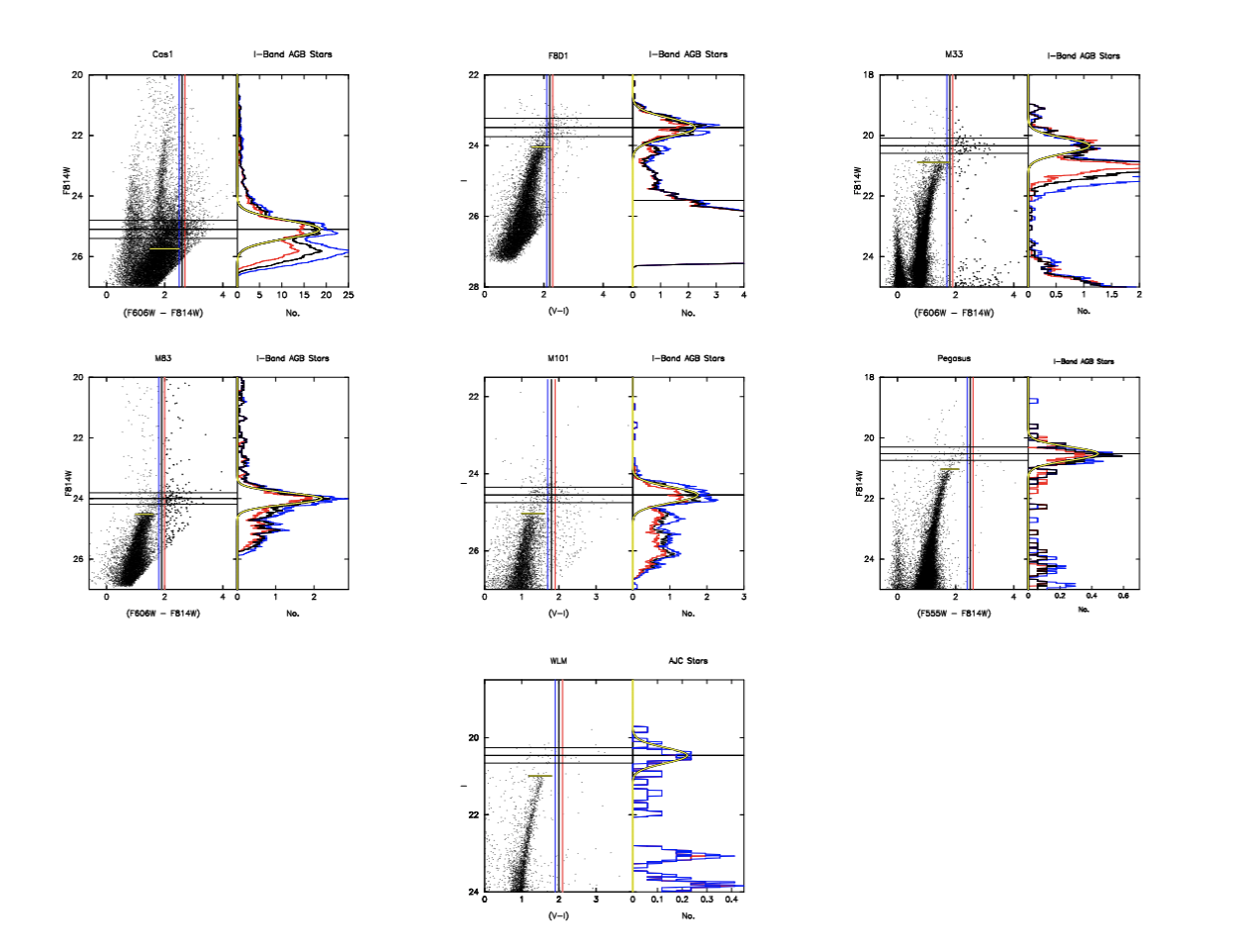}
\caption{Named Galaxies. See Figure 3 for description.}
\end{figure}

\begin{figure}
\centering
\includegraphics[width=18.0cm, angle=-0]{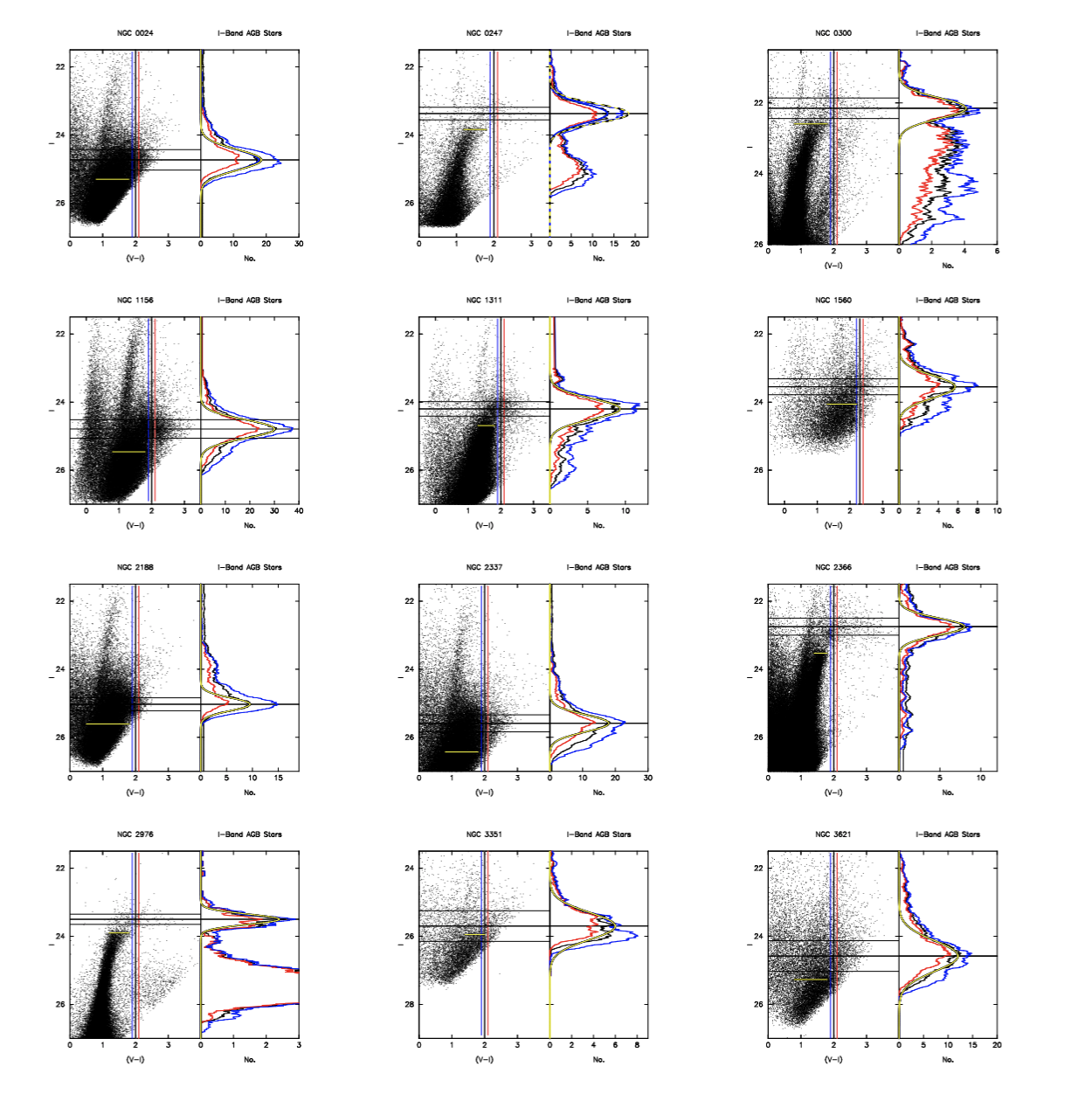}
\caption{NGC galaxies. See Figure 3 for description.}
\end{figure}

\begin{figure}
\centering
\includegraphics[width=18.0cm, angle=-0]{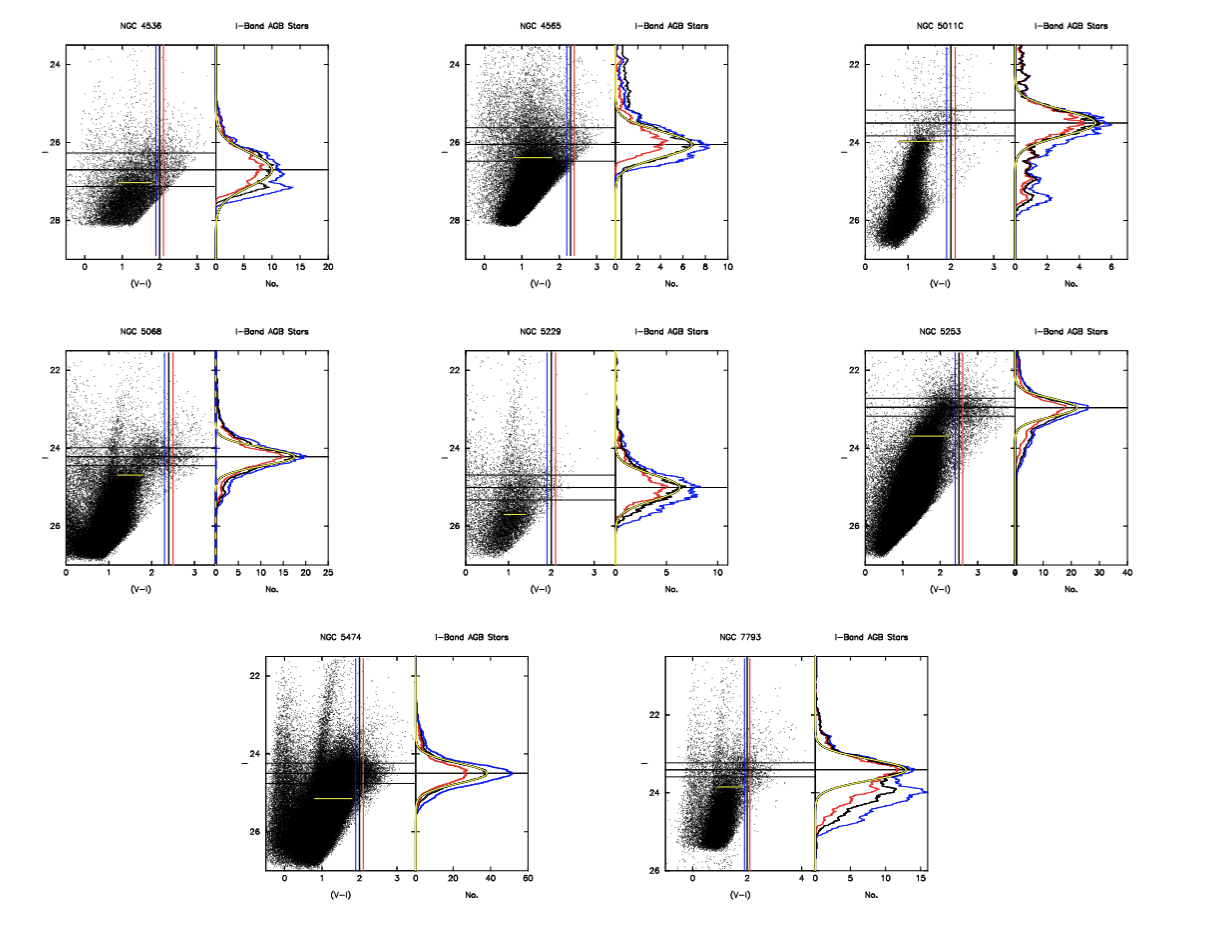}
\caption{NGC galaxies (cont.) See Figure 3 for description.}
\end{figure}


\begin{figure}
\centering
\includegraphics[width=18.0cm, angle=-0]{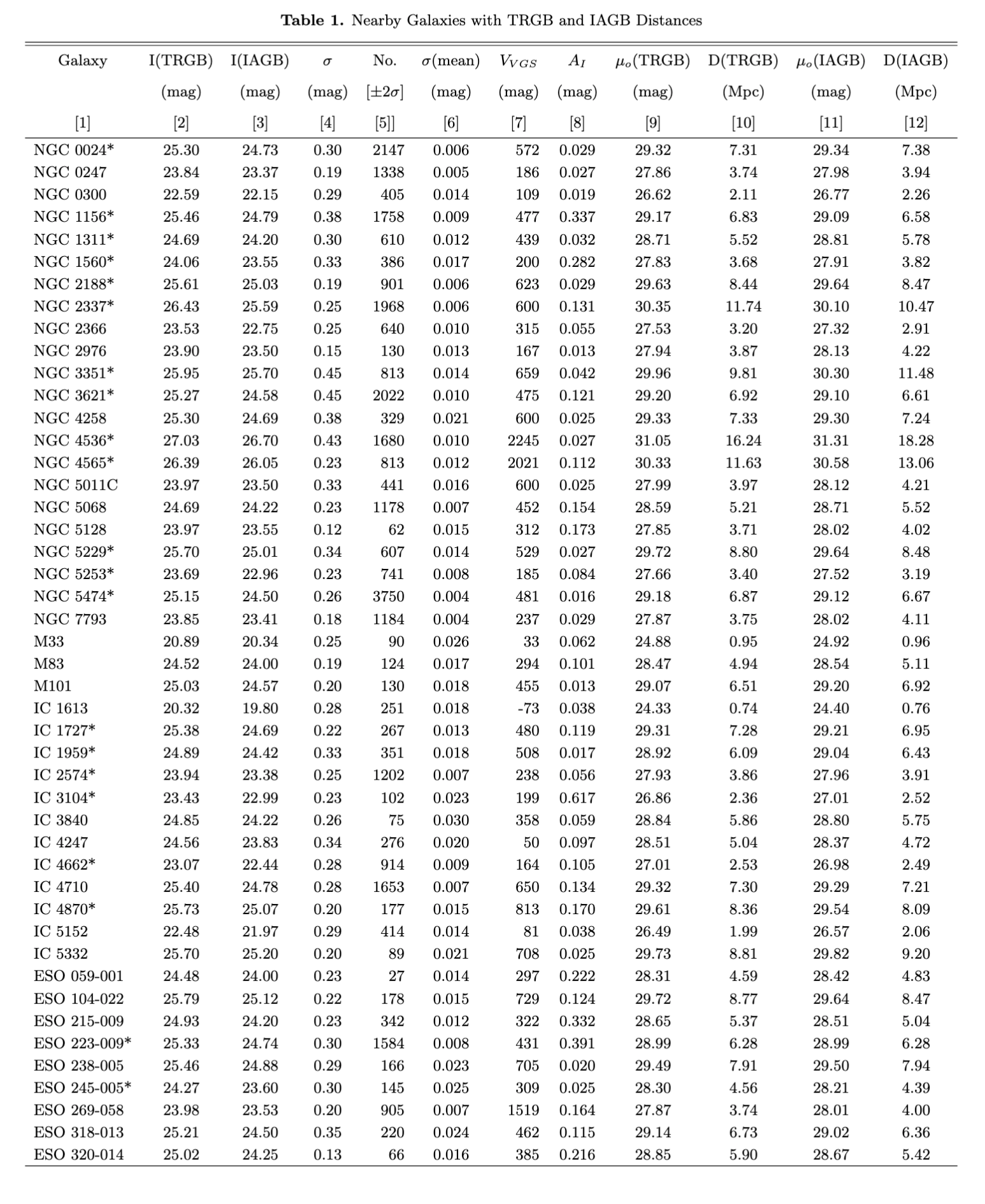}
\end{figure}

\begin{figure}
\centering
\includegraphics[width=18.0cm, angle=-0]{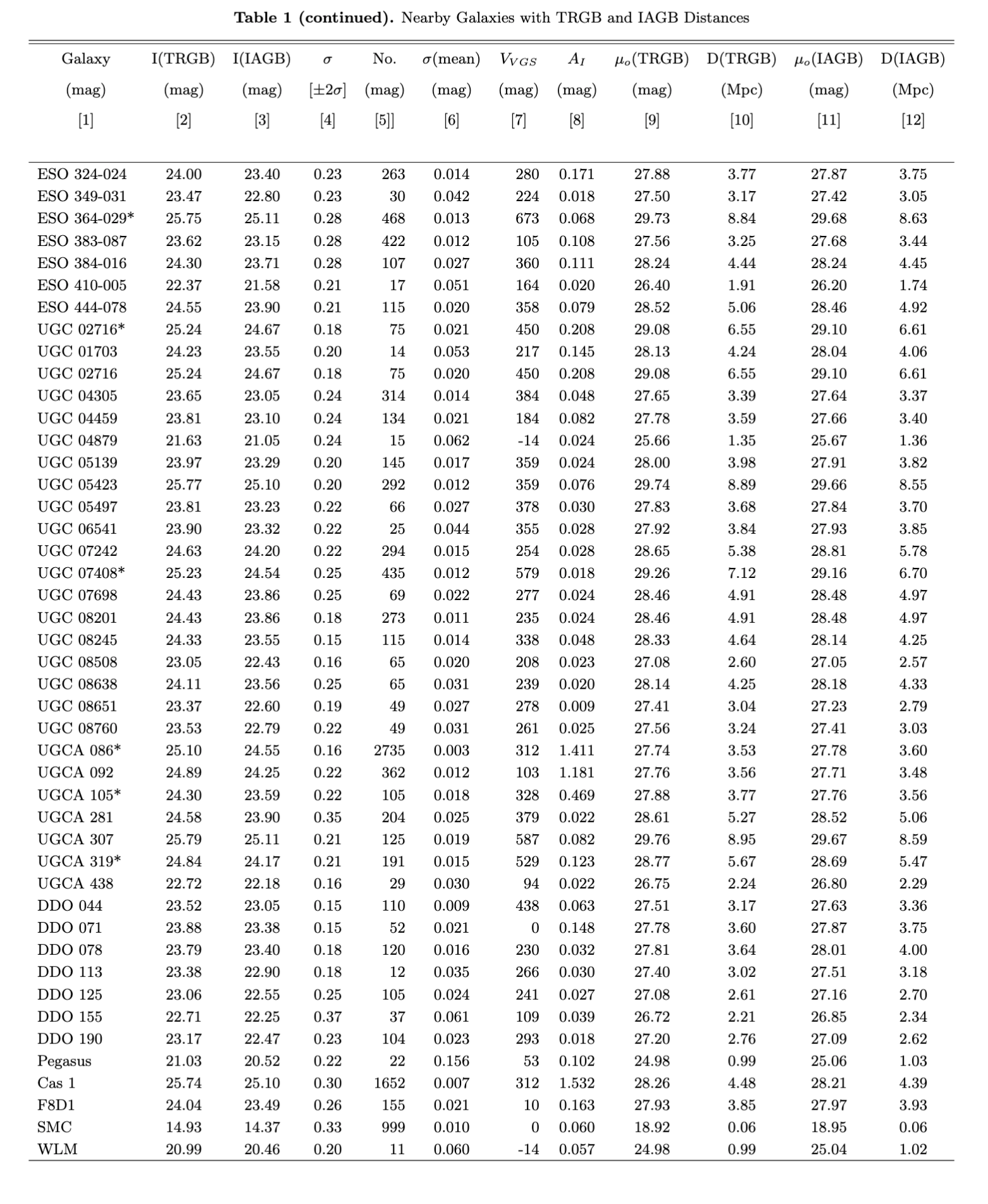}
\end{figure}

\clearpage
\subsection{Color Magnitude Diagrams}\label{sec:sec4}

Figures 3 through 14 contain the color-magnitude diagrams to the left within each sub-panel, for the galaxies with {\it IAGB} distances derived in this paper. The right side of each sub-panel contains the marginalized luminosity function (oscillating black line) for those stars with colors in excess of the vertical black line shown in the CMD. The blue curve is a Gaussian fit in height and width to the core of the luminosity function. Red lines to either side of the adopted black luminosity function are the luminosity functions resulting from perturbing the color cut by $\pm$0.10~mag, given to illustrate the sensitivity of the luminosity function to  changes in the color selection function.  
\clearpage

\section{Summary and Conclusions}\label{sec:sec6}

Following the introduction and zero-point calibration of a new method of determining extragalactic distances (Madore et al. 2024) using I-band Asymptotic Giant Branch {\it IAGB} stars, we have undertaken an exploratory extended application out to 9~Mpc. I vs (V-I) color-magnitude diagrams for 92 galaxies, have been used to determine (a) the apparent I-band (F814W) magnitude of the TRGB and (b) within the same diagram, for the same galaxies, determine the I-band modal magnitude of the very reddest population of high-luminosity stars, the {\it IAGB} stars. From this comparison it is found that the {\it IAGB} stars are -0.589~mag brighter than the TRGB, where the standard error on that difference is $\pm 0.119$~mag. Dividing the contributions to this cumulative error equally between the two distance indicators suggests that they would both have intrinsic dispersions of $\pm$ 0.084~mag in that case. This is probably high for the TRGB, suggesting that a more equitable estimate might be $\pm$0.06~mag for the {\it TRGB} and $\pm$0.10~mag for the {\it IAGB} at this early stage in its application. { The effect of differences in the star formation	histories, differences
in the mean metallicities of the AGB populations, and differences in
the {\it in situ} reddenenings between galaxies in this sample must be
contained within the 0.10 mag scatter cited above, as  found in the
comparison of TRGB and IAGB distance moduli.
}

The total error on the intercomparison yields an error on the mean between the two distance indicators of $\pm$ 0.012~mag.
This relative distance zero point can be used to provide an independent calibration of the {\it IAGB} distance determination method. Adopting $M_I(TRGB) = $ -4.05~mag for the tip of the red giant branch stars gives $M_I(IAGB) = $ -4.64~mag $\pm$ 0.012 (stat) $\pm$ 0.02 (sys). 

We {\bf close by noting} that the data used for this study were not taken with the purpose of measuring {\it IAGB} distances, and that further targeted observations taken with consistent instrumentation and calibrations will improve future applications. Thus, with single-epoch observations (using HST and/or Roman), it is reasonable to expect that these stars will be capable of independently measuring the distances to the majority of the Type Ia supernova calibrating galaxies (i.e., within 40~Mpc) currently forming the basis for determining the local value of the expansion rate of the universe.  Using NIRCam and its F090W filter, JWST should be able to apply the method even further. 

\acknowledgments
The data presented in this paper were obtained from the Extragalactic Distance Database (EDD). We thank the {\it University of Chicago} and {\it the Observatories of the Carnegie Institution for Science} for their support of our continuing research into the calibration and determination of the expansion rate of the Universe. 

\facility{HST (ACS/WFC, WFC3/IR)}

\section{References}
\par\noindent
Freedman, W.F. 2021, ApJ, 919, 16
\par\noindent
Freedman, W.F. \& Madore, B.F. 2023, JCAP, 11, 50, arXiv.2309.05618
\par\noindent
Freedman, W.~L., Madore, B.~F., Hoyt, T., et al.\  2020, \apj, 891, 57
\par\noindent
Hoyt, T.J., 2023, NatAs, 7 590
\par\noindent
Graczyk, D., Pietrzy{\'n}ski, G., Thompson, I.~B., et al.\ 2014, \apj, 780, 59. 
\par\noindent
Lee, A. 2023, ApJ, 956, 15
\par\noindent
Madore, B.F., Freedman, W,L., Hoyt, T.J., et al. 2025, ApJ, in press, (Paper I)
\par\noindent
Pietrzy{\'n}ski, G., Graczyk, D., Gallenne, A., et al.\ 2019, \nat, 567, 200. 
\par\noindent
Reid, M.~J., Pesce, D.~W., \& Riess, A.~G.\ 2019, \apjl, 886, L27 

\par\noindent
Skowron et al. 2022

\par\noindent
Scowcroft, V., Freedman, W.~L., Madore, B.~F., et al.\ 2016, \apj, 816, 49. 
\par\noindent
Wu, J., Scolnic, D., Riess, A.G., et al.\ 2023, \apj, 954, 87

\vfill\eject
\begin{figure*}
\centering
\includegraphics[width=1.2\textwidth]{"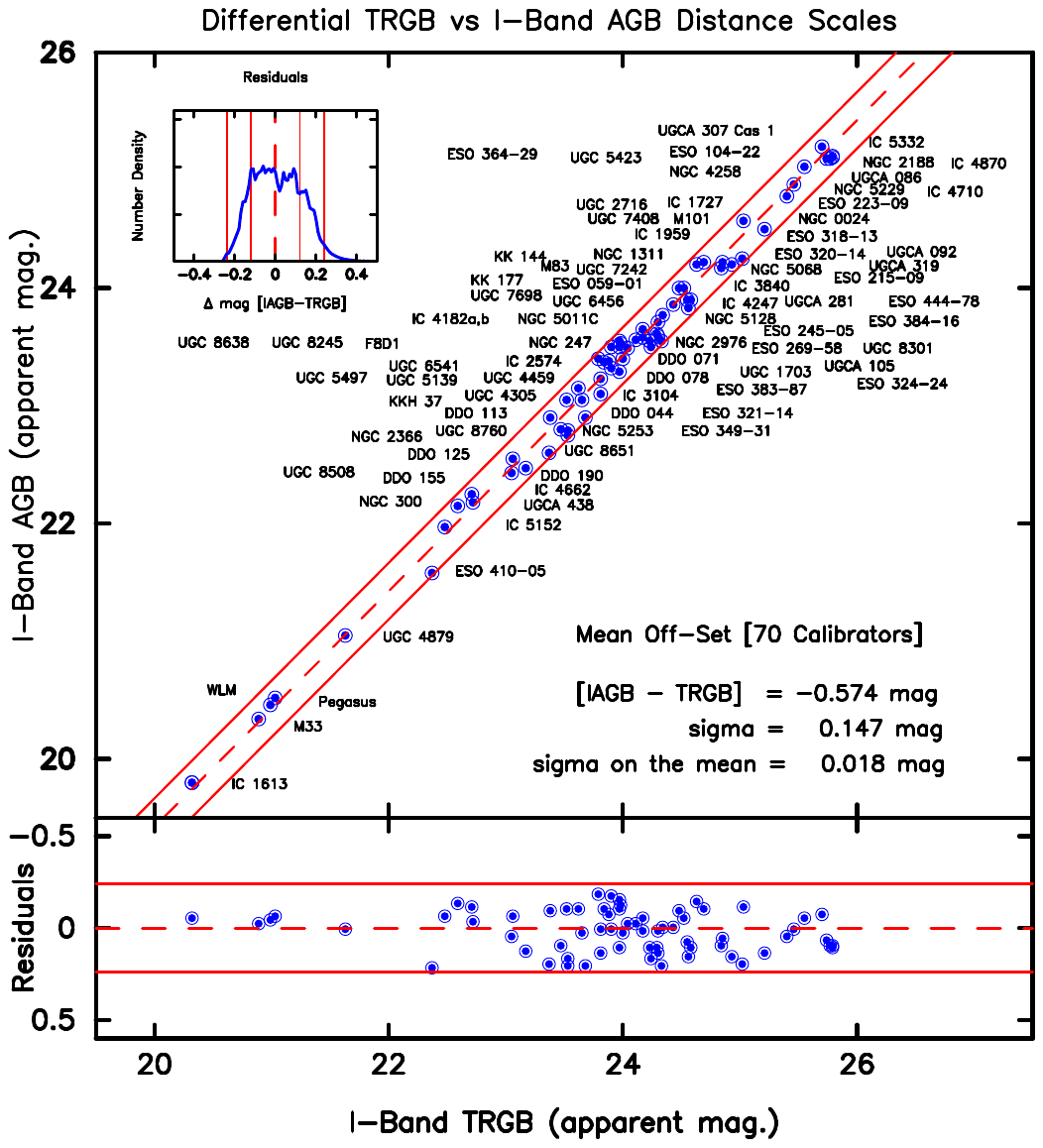"}
\caption{Same as Figure 1 except this sample has been down-selected to remove 23 galaxies that were deemed to be sub-optimal for TRGB measurements based on the placement of their points largely in the disks if these galaxies. }
\end{figure*}
\clearpage
\section{Appendix}
\subsection{Analysis of a Down-Selected Subsample of 70 Galaxies}
A number of the galaxies discussed in the main paper above had their original pointing concentrated on the disks of the target galaxies and, they are therefor less than optimal for detecting and measuring the halo TRGB population. 
The names of those 22 galaxies are marked with a following asterisk in Table 1, and an inspection of their CMDs bear witness to the significant presence of Population I stars in the form of blue and red supergiants. 

Figure 16 shows the resulting correlation of I-Band AGB distances versus I-Band TRGB distances for the remaining 70 down-selected galaxies. The quantitative changes in the correlation are small, and perhaps slightly counter-intuitive. The mean difference [IAGB-TRGB] has become less negative (-0.574~mag for the smaller sample as compared to -0.589~mag for the larger sample) suggesting that the TRGB population superimposed on the disks that were dropped from the sample, were (as might have been anticipated) biasing the solution towards brighter magnitudes. On the other hand, drawing down the bias by rejecting the (potentially) crowded  TRGB sample results in a net increase in the scatter around the mean, with it going from $\pm$0.119~mag up to $\pm$ 0.147~mag. One might have predicted that the scatter would decrease in removing a biased subset of the data, but it did not.

\end{document}